\documentclass[12pt,a4paper]{article}
\usepackage[utf8x]{inputenc}
\usepackage{ucs}
\usepackage{graphicx}
\usepackage{geometry}
\begin{document}
\title{Numerical investigation of the electrical conductivity of irradiated graphene}
\author{D.V.Kolesnikov$^1$\\$^1$\small{Bogoliubov Laboratory of Theoretical Physics,}\\
\small{Joint Institute for Nuclear Research (JINR), Dubna, Russia}\\
\small{e-mail:kolesnik@theor.jinr.ru}}

\maketitle
\abstract {Transport properties of irradiated graphene (electrical conductivity and mobility) are numerically investigated using the real-space Kubo formalism. A micrometer-sized system consisting of millions of atoms with nanopores of various sizes and concentrations is described. Electrical conductivity and mobility as a function of carrier (hole) density are calculated to provide possible comparisons with experiments.}
\section{Introduction}
The investigation of electrical conductivity and mobility in the defect samples of graphene and other novel 2D materials is an important and challenging task of the modern condensed matter science. While the pristine samples of 2D materials shows high values of electron mobility, several types of mechanisms which can possibly decrease it's value exist. Among these are the electron-phonon interaction \cite{e-ph1}, point-like defects (vacancies, Stone-Wallace defects, charged impurities) \cite{point1,point2} and the distributed defects and defect formations, such as domain walls and disclination dipoles \cite{dd1}. On the other hand, one should also mention the defect engineering approach~\cite{de1}, when defect are artificially introduced into the sample  to tailor for desired properties. The most direct approach of altering the electronic properties of graphene is doping~\cite{dope1,dope2}. The downside of this method is the degradation of the sample quality \cite{deg1}, which limits the usage of chemically modified graphene in electronics applications. Other methods, that does not involve the chemical modification may include the use of electron beams, laser pulses or particle irradiation~\cite{beam1,beam4}. Such modifications may introduce various types of defects into graphene \cite{beam2,beam3}, which can be described as nanopores, antidots or antidot arrays. One should note, that various estimations for the types and sizes of radiation-induced defects exist. The defect types may vary from single vacancies and Stone-Wallace defects, to the large nanopores, possibly with additional edge structures \cite{beam2, irrad1}. The structure of the nanopores edge may lead to the presence of edge electron states at the low energies, which may alter electronic and transport properties. While the periodic antidots can be described with the well-known approaches such as the non-equilibrium Green's function \cite{negf1}, structures with random distribution of defects requires a different method and/or larger size of the supercell. Such structures can be synthesized using both top-down and bottom-up \cite{pore-exp} approaches.\quad

Several theoretical approaches can be used to describe electron transport properties of the defect graphene. Various empirical approaches, based on the variations of Boltzmann kinetic equation, were developed \cite{bo1,bo2}. On the other hand, the first principles approach based on the Landauer equation and the non-equilibrium Green's function (NEGF) method \cite{negf1,negf2} take into account many rich phenomena of electron and transport properties of defect graphene. The downside of this approach is the significant limitation of the atom number of the sample or the supercell. Thus the NEGF method is used for small samples, scattering processes with low correlation lengths, or in the ballistic transport regime. To utilise the first-principles approach for the micrometer-sized defect graphene samples, effectively determining the conductivity and mobility in the diffusive regime, the real-space Green-Kubo method is  used\cite{GK1,GK2}. A combination of the Green-Kubo or Einstein formula for the conductivity\cite{KG-Gr,lstm} with the kernel polynomial method \cite{KPM} allows effective estimation of the conductivity for the randomly distributed patterns in graphene \cite{triangles}.  We take into account the effects of irradiation by introducing randomly distributed nanopores of fixed radius into the sample. We introduce periodical boundary conditions to avoid problems arising from the finite sample size. We find the mean square deviation (MSD) of the electron as a function of the correlation time with the use of kernel polynomial method. From the MSD, we find the electrical conductivity with the Einstein's formula.

\section{Computational formalism}
We start from the Green-Kubo formula for the zero-temperature electrical conductivity
\begin{equation}
\sigma^{GK}_{\mu\nu}(E)=\frac{2\pi\hbar e^2}{\Omega} Tr[\hat{V}_\mu \delta(E-H)\hat{V}_\nu \delta(E-H)],
\end{equation}
where $\sigma^{GK}$ is the conductivity, $\hbar$ being the reduced Planck constant, $e$ is the electron charge, 
$\Omega$ is the 2D volume,$\delta$ represents the delta function, $E$ is the energy, $H$ is the Hamiltonian of the system and $\hat V_\mu, \mu=x,y$ is the velocity operator in the $\mu$ direction. Note that the factor 2 in $2\pi\hbar e^2/\Omega$ represents the spin degeneracy of the Hamiltonian. One can use the Fourier transformation of the delta-function $\delta (E-H) = \int_{-\infty}^\infty \exp(i(E-H)t/ \hbar) dt/{2\pi\hbar},$ so that
the conductivity is found as 
\begin{equation}
\sigma^{GK} = \frac{e^2}{\Omega}\int_{-\infty}^\infty dt Tr[e^{iHt}\hat{V} e^{-iHt}\hat{V}\delta(E-H)],
\end{equation}
where $\hat{V}$ is the velocity operator in x direction. Using the formulas above, one can find the Green-Kubo formula for the running electrical conductivity $\sigma^{GK}(E,t)$ and the density of states (DOS) $\rho(E)$ from the velocity autocorrelation function $C_{vv}$ as
\begin{eqnarray}
\sigma^{GK} (E,t)=e^2 \rho(E) \int_0^t C_{vv}(E,t) dt,\\
C_{vv}(E,t) = \frac{Tr[\frac{2}{\Omega} \delta(E-H)(\hat{V}(t)\hat{V} + \hat{V}\hat{V}(t))/2]}{Tr[\frac{2}{\Omega}\delta(E-H)]},\\
\rho(E) = Tr[\frac{2}{\Omega}\delta(E-H)],
\end{eqnarray} 

where $\hat{V}(t) = \hat{U}^\dagger(t)\hat{V}\hat{U}(t) = e^{iHt/\hbar}\hat{V}e^{-iHt/\hbar}$ is the velocity operator in the Heisenberg representation, and $\rho(E)$ is the density of states. Furthermore, by integrating the Green-Kubo formula, one can 
find the Roche-Mayou formula \cite{RoMa} for the running conductivity as a derivative of the MSD $\Delta X^2(E,t)$
\begin{eqnarray}
\sigma^E(E,t) = e^2 \rho(E) \frac{1}{2}\frac{d}{dt}\Delta X^2(E,t),\label{sigmaE}\\
\Delta X^2(E,t) = \frac{Tr[\frac{2}{\Omega}\delta(E-H)(\hat{X}(t)-\hat{X})^2]}{Tr[\frac{2}{\Omega}\delta(E-H)]},
\end{eqnarray}
where $\hat{X}(t)=\hat{U}^\dagger (t)\hat{X}\hat{U}(t)$ is the x-coordinate operator in the Heisenberg representation. To estimate the conductivity from the running conductivity, we use the large time limit $\sigma(E)=\lim_{t\rightarrow \infty}\sigma^{E}(E,t)$. 
One should also note, that instead of (\ref{sigmaE}) the formula $\sigma=e^2\rho \Delta X^2/(2t)$ is often used. However, we found that the usage of this formula leads to the significant overestimation of the conductivity in comparison with (\ref{sigmaE}) for the correlation times t$\approx$80-100 ns used in our calculation. To ensure the validity of the approximation, we plot the value of REC as a function of time for different energies, showing the convergence of REC. One should also note, that one should consider the question of the transport regime in the sample. The actual transport regime depends on the values of the elastic mean free path and the localization length~\cite{ZBW}, as well as the sample size. We assume the diffuse dominant transport regime within our calculation.\quad

\begin{figure}[h!]
\begin{center}
\includegraphics[width=7cm]{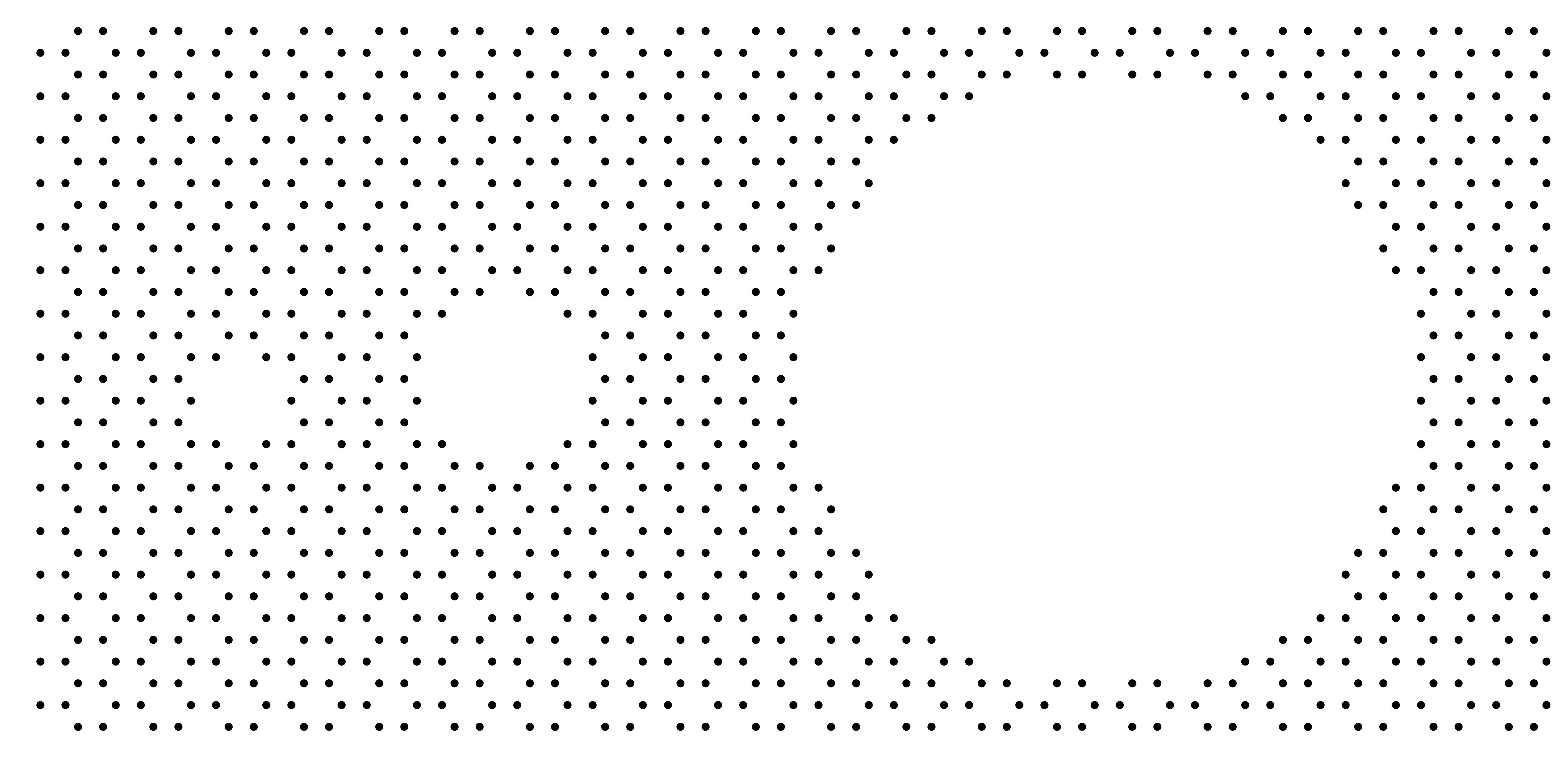} 
\caption{Three types of the nanopore defects considered: r= 1.5 \AA~  nanopore with the single hexagonal ring removed (left), r=3 \AA~  nanopore (center) and r=12 \AA~ nanopore (right).}
\label{fig1}
\end{center}
\end{figure}

For the numerical calculation of the conductivity from the Einstein formula (\ref{sigmaE}), the kernel polynomial method is used~\cite{KPM}. First, we rescale the Hamiltonian, energy and time as $H\rightarrow H/\Delta E, E\rightarrow E/\Delta E, t\rightarrow t \Delta E$, where $\Delta E$  is the scaling energy parameter, so that the eigenvalues of scaled Hamiltonian lies within the interval [-1,1]. By expanding the delta function in (\ref{sigmaE}) by the Chebyshev polynomials $T_n(E), n=0\ldots N_m-1$,  one can find 
\begin{eqnarray}
\rho(E) = \frac{2}{\pi\Omega\Delta E\sqrt{1-E^2}}\sum_{n=0}^{N_m-1}g_n(2-\delta_{n0})T_n(E)C_n^{DOS},\label{rhoE}\\
\rho(E)\Delta X^2(E,t) = \frac{2}{\pi\Omega\Delta E\sqrt{1-E^2}}\sum_{n=0}^{N_m-1}g_n(2-\delta_{n0})T_n(E)C_n^{MSD}(t),
\end{eqnarray}
where $g_n=(1-n\alpha)\cos (\pi n \alpha) + \alpha \sin (\pi n \alpha)\cot (\pi\alpha)$ is the Jackson damping function, $\alpha = 1/(N_m + 1)$, $\delta_{ij}$ is the Kronecker symbol, $T_n(E)$ is the nth Chebyshev polynomial, and $C_n^{DOS}, C_n^{MSD}(t)$ are the Chebyshev moments. The moments are found using the kernel polynomial method \cite{KPM},  using $N_r$ random vectors $|\phi>$
\begin{eqnarray}
C_n^{DOS} \approx <\phi|T_n(H)|\phi>,\label{Cn}\\
C_n^{MSD} \approx <\phi|[\hat{X},\hat{U}(t)]^\dagger T_n(H)\hat{U}^\dagger [\hat{X},\hat{U}(t)]|\phi>\label{Cmsd}.
\end{eqnarray} 
By using (\ref{rhoE}), (\ref{sigmaE}) and (\ref{Cn}), (\ref{Cmsd}) the density of states and conductivity are estimated. To calculate the correlators $[\hat{U}(t),X]$, the recursive algorithm is used (see \cite{KG-Gr, GPUQT}). Note, that by limiting the 
delta function expansion with $N_m$ Chebyshev polynomials we achieve the finite resolution by energy, determined as $dE\approx\pi\Delta E/N_m$. As for the estimation of the relative error due to the finite number of random vectors, it decreases with  $N_r$ increasing as $1/\sqrt{N N_R}$, where $N$ is the dimension of the Hamiltonian. Therefore, the real-space Green-Kubo method is especially useful for large systems. We have used the simple graphene nearest-neighbour tight-binding Hamiltonian $H = t\sum_{<mn>}\left|m\right\rangle\left\langle n\right|$, $t$=2.7 eV. To estimate the conductivity of the beam-irradiated graphene, we use the 2 million (2000 to 1000) atoms graphene sample with periodic boundary conditions in both x and y directions. We remove the certain number of randomly distributed circular patterns from graphene with radiuses equal to $\approx$ 1.5 , 3 and 12 angstrom (see Fig.\ref{fig1}), so that the fixed number of atoms (1 to 8 percent of the total) is removed. 

We calculate the density of states and the electrical conductiviy from the real-space Green-Kubo formalism using (\ref{rhoE}) and (\ref{sigmaE}), with the use of the kernel polynomial method.  For the density of states, we use $N_m$ = 1800 polynomials (with the corresponding energy resolution 0.01 eV) and $N_r$=64 random vectors, while for the computationally expensive estimation of conductivity the parameters are $N_m=800,\; N_r=64$. The MSD was fond to behave linearly for all the defect concentrations at $t\approx 60-100$ ns, and the electrical conductivity was estimated from the calculation of the time derivative of MSD in (\ref{sigmaE}) within this interval. To provide some comparison with the experiments, where the conductivity of the defect graphene is measured from the current-voltage characteristics, the hole carrier density was calculated for T=300 K from the density of states as
\begin{equation}
  n_s(E_F) = \int_{E_D}^\infty \rho(E)f(E-E_F)dE,\label{ns}
\end{equation}  
where $E_D$ is the graphene Dirac point (equal to zero energy in our case) and f is the Fermi-Dirac distribution function. Using the equation above, the conductivity as a function of the carrier density was found. One can also express the conductibvity as a function of the mobility $\mu$ as $\sigma=e n_s\mu$, to find the mobility in the form $\mu(E)=\sigma(E)/(e n_s)$.
\section{Numerical results}
\begin{figure*}[h!]
\begin{center}
\includegraphics[width=14cm]{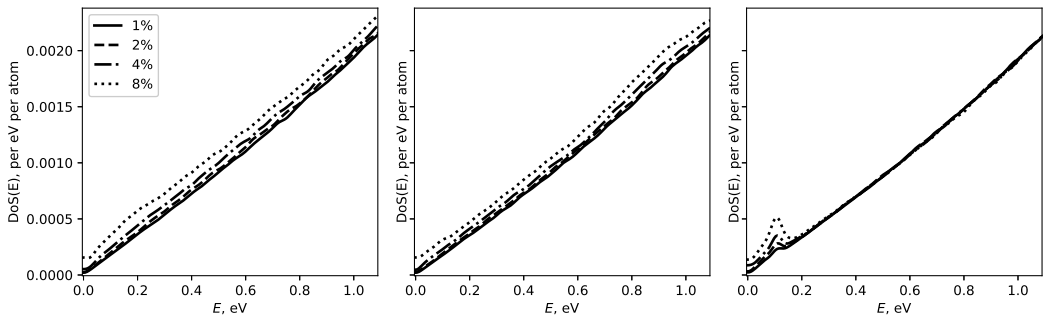} 
\end{center}
\caption{The density of states per eV per atom, as a function of energy. The cases of r=1.5\AA~ (left), 3\AA~ (center) and 12\AA~  nanopore defects (right) are shown. The solid, dashed, dash-dot and dotted lines represent 1\%, 2\%, 4\% and 8\% defect vacancy rate.}
\label{fig2}
\end{figure*}
The density of states (per eV per atom) for the irradiated graphene models described above is shown in Fig.\ref{fig2}. One can see the slight increase of density of states for 1.5\AA~ and 3\AA~ nanopores at the Fermi energy (E=0), increasing with nanopore concentration, and the appearance of defect peaks in the density of states for 12\AA~ nanopores at the low energies about 0.1 eV. The density of states, aside from the peaks, shows linear dependence on the energy, which is typical for pristine graphene. Note, that the overall density of states \textit{per atom}, shown in Fig.\ref{fig2}, is generally independent on the concentration of defects or may even show slight increase for the smaller nanopore types, while the density of states \textit{per area} $\rho(E)$ (see (\ref{rhoE})~) is reduced proportionally to the defect concentration, thus it decreases with the concentration of defects increasing. 

\begin{figure*}[h!]
\begin{center}
\includegraphics[width=14cm]{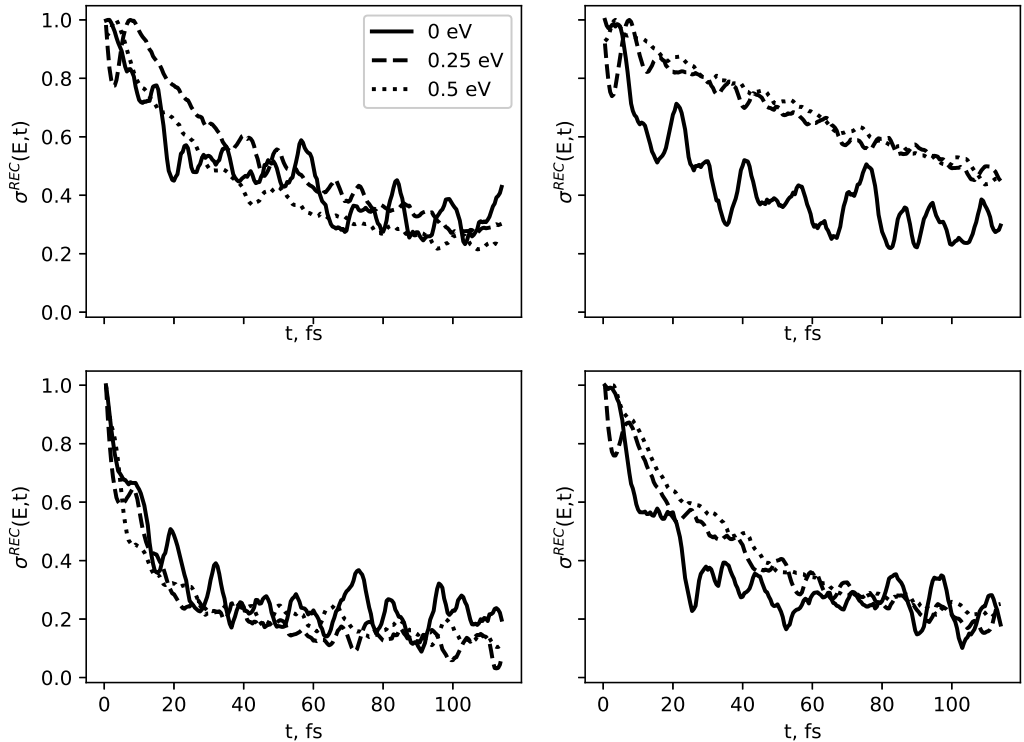} 
\end{center}
\caption{The normalized running electric conductivity (REC) as a function of energy and time (in fs), for r=1.5\AA~ (left column) and r=12\AA (bottom column), with defect concentrations 1 percent (top row) and 8 percent (bottom row).}
\label{fig2a}
\end{figure*}
The running electric conductivity (REC) as a function of time is presented in Fig.~\ref{fig2a}. One can see the REC convergence for times t$\geq$ 100 fs for all the cases considered, which is more pronounced for smaller nanopores (left column) and higher energies. The low-energy REC ($E=0$) is fluctuating strongly with time, which can be attributed to the Zitterbewegung phenomenon \cite{ZBW}.

\begin{figure*}[h!]
\begin{center}
\includegraphics[width=14cm]{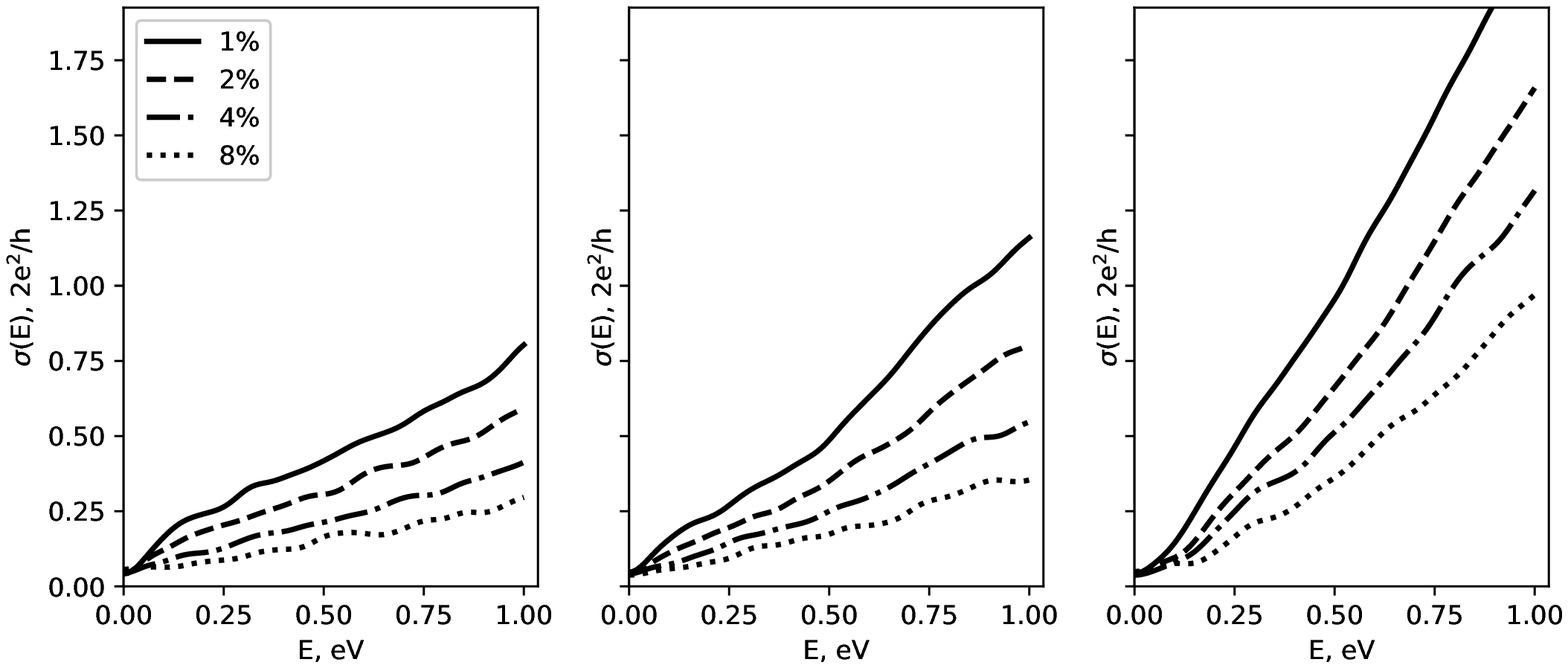} 
\end{center}
\caption{The electrical conductivity as a function of energy, found from the Einstein's formula (\ref{sigmaE}). The cases of r=1.5\AA~ (left), 3\AA~ (center) and 12\AA~  nanopore defects (right) are shown.}
\label{fig3}
\end{figure*}

The electrical conductivity as a function of energy, in the units of $2e^2/h$, is shown in Fig.\ref{fig3}. One can see, that the conductivity decreases with the density of defects increasing and defect radius decreasing. At the charge neutrality point $E_F=0$, a finite value of  conductivity is present for all the cases considered. One can also see the non-monotonic behavior of conductivity, which is more pronounced for maximum concentration of defects (8 \%). Note that the presence of defect peaks in the density of states (see Fig.\ref{fig2}) does not have significant influence on the conductivity. One should also note the dependence of the conductivity on the nanopore concentration: for the 1.5\AA~ and 3\AA~ nanopores, the twofold increase of the defect concentration leads to the approximately 25 \% decrease of the conductivity.

\begin{figure*}[h!]
\begin{center}
\includegraphics[width=14cm]{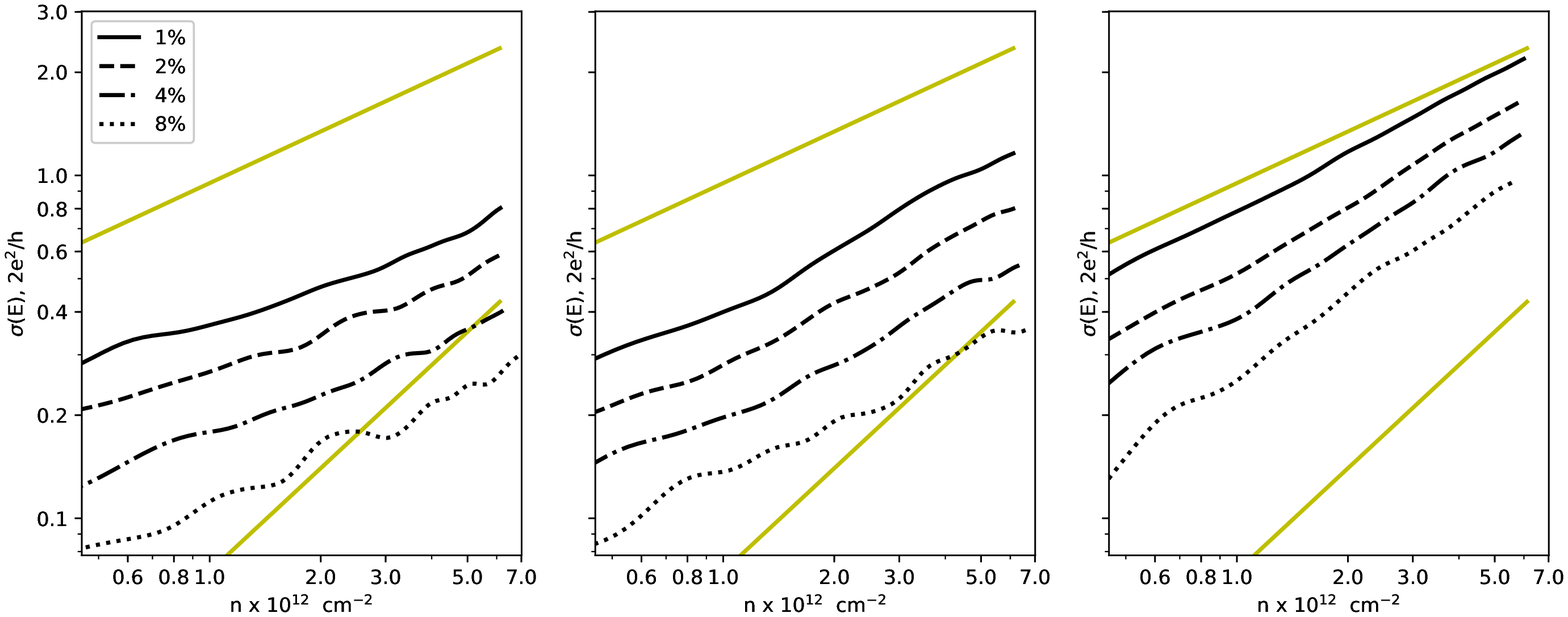} 
\end{center}
\caption{The conductivity as a function of the carrier (hole) concentration $n_s$ (see \ref{ns}). The cases of r=1.5\AA~ (left), 3\AA~ (center) and 12\AA~ nanopore defects (right) are shown. The lines represent the functions $f\sim n^{1/2}$ and $f\sim n$.}
\label{fig4}
\end{figure*}

The electrical conductivity as a function of the carrier (hole) density $n_s$, is shown in Fig. \ref{fig4}. One can see, that the conductivity generally follows the $\sigma\sim n_s^\alpha$ law, where $\alpha\approx 1/2$, which is typical behaviour for the ballistic scattering \cite{RadS}. For the smaller nanopores, one can estimate $\alpha\leq 1/2$, while for the larger 12\AA~  nanopore the $\alpha$ could have larger value $\alpha\geq 1/2$.  
\begin{figure*}[h!]
\begin{center}
\includegraphics[width=14cm]{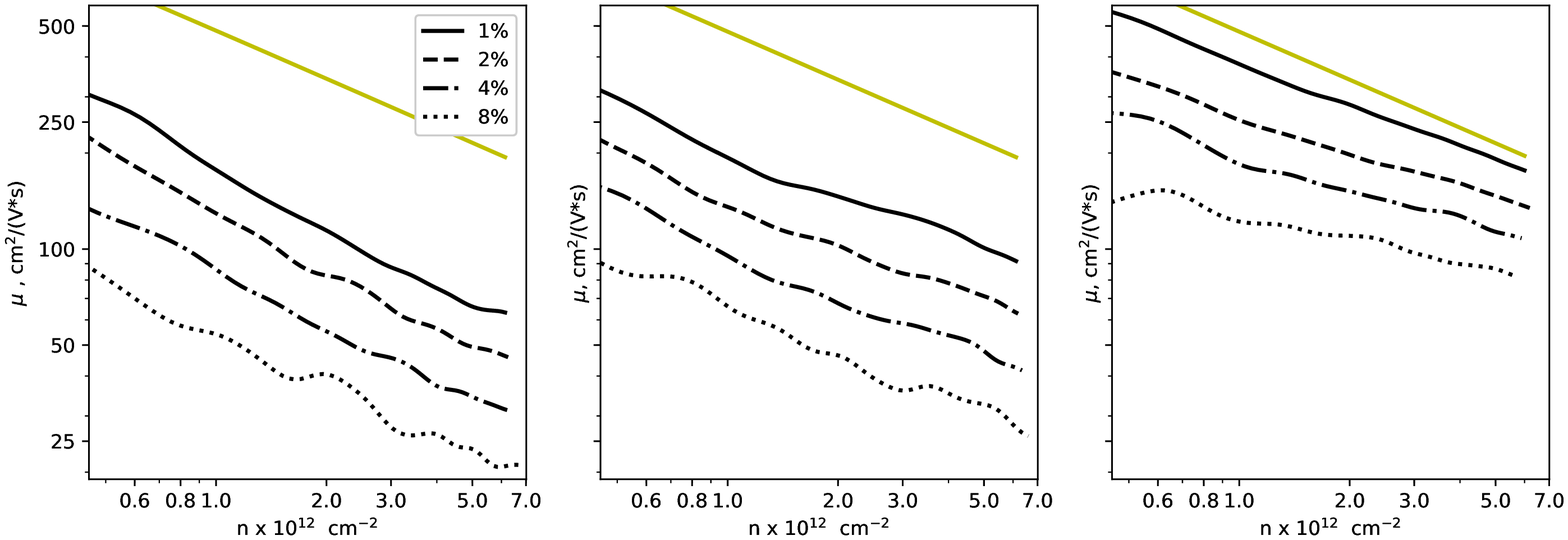} 
\end{center}
\caption{The mobility $\mu$ as a function of the carrier (hole) concentration $n_s$ (see \ref{ns}). The cases of r=1.5\AA~ (left), 3\AA~ (center) and 12\AA~ nanopore defects (right) are shown. The line represents the function $f\sim n^{-1/2}$.}
\label{fig5}
\end{figure*}
The electron mobility as a function of the carrier density, is shown in Fig.\ref{fig5} One can see the decrease of the mobility with the carrier density increasing, nanopore concentration increasing and the nanopore radius decreasing. The overall behavior of the mobility can be estimated with the formula $\mu\sim n^\beta$, where $\beta\leq -1/2$ for the smaller 1.5\AA~ and 3\AA~ nanopores and $\beta\geq -1/2$ for the larger 12\AA~ nanopore. 

\section{Conclusion}
We have investigated the electrical conductivity and electron mobility for the graphene sample with randomly distributed circular nanopores. We have found the density of electronic states, the conductivity and mobility as a function of energy, nanopore concentration and nanopore radius. The behaviour of the density of states per atom is found to be close to the linear behaviour of pristine graphene sample. For the smaller type of nanopores with radius less than 3\AA~, the DOS is increased near the Fermi energy, while for the larger nanopores the small peak of DOS near E$\approx$0.1 eV is observed.  The conductivity is found to increase with energy increasing and the defect vacancy rate decreasing. The twofold increase of the vacancy rate leads to the approximately 25 \% decrease of the conductivity, which is more pronounced for smaller nanopores.  For the samples with different nanopore radius and the same vacancy concentration, the conductivity is increased with nanopore radius increasing. The conductivity as a function of the carrier concentration $n$ was found, and it shows the $ n^{1/2}$ behaviour, while the mobility shows $n^{-1/2}$ behaviour. For the 3\AA~ and smaller nanopore types, the conductivity is increased slightly slower than $n^{1/2}$, while the 12\AA~ nanopore shows faster increase with energy.

\end{document}